\documentclass{PoS}
\usepackage{subfigure}
\usepackage{amsmath}
\usepackage{amssymb}
\usepackage{bbm}

\newcommand{\MSb}{\overline{\textrm{MS}}}

\newcommand{\chibar}{\overline{\chi}}

\newcommand{\expect}[1]{\langle{#1}\rangle}

\newcommand{\PM}{\mathbbm{P}_M}
\newcommand{\RM}{\mathbbm{R}_M}
\def\dirac#1{\gamma_{#1}}
\def\Tr{{\rm Tr}}

\title{{\vspace{-15mm} \normalsize\hfill{\small DESY 13-207}}\\[15mm]
{\vspace{-18mm}\normalsize\hfill{\small HU-EP-13/63}}\\[12mm] 
{\vspace{-15mm}\normalsize\hfill{\small SFB/CPP-13-95}}\\[10mm] Topological susceptibility from twisted
mass fermions using spectral projectors}
 
\ShortTitle{Topological susceptibility from twisted mass fermions using spectral projectors}

\author{\speaker{K.~Cichy}$^{ab}$, E.~Garcia-Ramos$^{ac}$, K.~Jansen$^a$ and
A.~Shindler$^{d}$
\\
\llap{$^a$}NIC, DESY Zeuthen, Platanenallee 6, 15738 Zeuthen, Germany\\
\llap{$^b$}Adam Mickiewicz University, Faculty
of Physics, Umultowska 85, 61-614 Poznan, Poland\\
\llap{$^c$}Humboldt Universit\"at zu Berlin, Newtonstr. 15, 12489
  Berlin, Germany\\
\llap{$^d$}IAS, IKP and JCHP, Forschungszentrum J\"ulich, 52428 J\"ulich, Germany\\
E-mail: \email{krzysztof.cichy@desy.de}, \email{elena.garcia.ramos@desy.de},
\email{karl.jansen@desy.de}, \email{a.shindler@fz-juelich.de}
}

\abstract{
We discuss the computation of the topological susceptibility using the method of spectral projectors and
dynamical twisted mass fermions.
We present our analysis concerning the $O(a)$-improvement of the topological susceptibility and
we show numerical results for $N_f=2$ and $N_f=2+1+1$ flavours, performing a study of the
quark mass dependence in terms of leading order chiral perturbation theory.
\begin{center}
\vspace*{1cm}
\includegraphics
[width=0.2\textwidth,angle=0]
{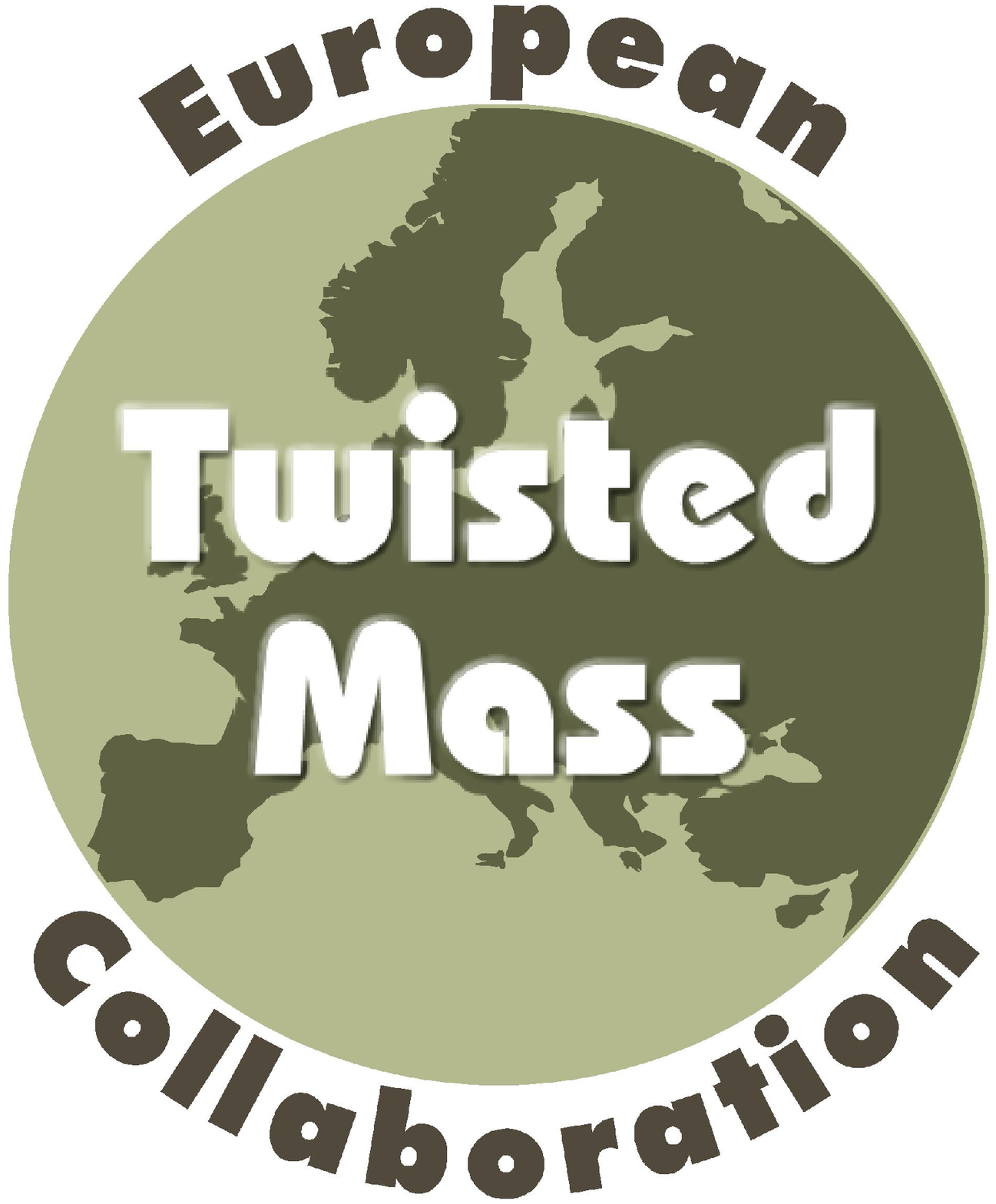}
\end{center}
}

\FullConference{31st International Symposium on Lattice Field Theory - LATTICE 2013\\
		July 29 - August 3, 2013\\
		Mainz, Germany}

\begin{document}

\section{Introduction}
The topological susceptibility, a quantity that expresses fluctuations of the topological charge of gauge
fields, can be linked to $n$-point correlation functions of sufficiently many scalar and
pseudoscalar quark densities, the so-called density chain correlation functions or density chains.
It was shown in Ref.~\cite{Giusti:2004qd,Luscher:2004fu} that density chains provide a definition of the
topological
susceptibility that is free of short-distance singularities and regularization-independent.
Moreover, this definition can be combined with the method of spectral projectors \cite{Giusti:2008vb} to
evaluate the topological susceptibility efficiently in terms of computing time.
The first application of spectral projectors to the quenched case was discussed in
Ref.~\cite{Luscher:2010ik} and our preliminary results for the dynamical case, using Wilson twisted mass
fermions, were shown in Ref.~\cite{Cichy:2011an}.
We refer to the upcoming publication for a more comprehensive discussion of our
results \cite{Cichy:2013chi}.

In these proceedings, we discuss the issue of $O(a)$-improvement of the
topological
susceptibility.
Twisted mass fermions are said to be automatically $O(a)$-improved at maximal twist
\cite{Frezzotti:2003ni}.
Specifically, this means that on-shell quantities that are $\mathcal{R}_5$-parity even (defined below)
can not have
$O(a)$ cut-off effects.
However, the topological susceptibility is defined via density chains that include 
integrals (sums) over all space time points leading to contact terms with short distance singularities.
The presence of contact terms can, in principle, spoil $O(a)$-improvement.
We use Operator Product Expansion (OPE) to show that this is not the case.
For a similar proof, concerning the improvement of the chiral condensate, we refer to
Ref.~\cite{Cichy:2013egr} in these proceedings.
After the discussion of the improvement, we also summarize some of our numerical results.

\section{Topological susceptibility from density chains}
In the continuum, the relation between the topological
charge $Q$ and density chain correlation functions can be
established via the equation $\textrm{Tr}\{\gamma_5f(D)\}=f(0)Q$,
where $D$ is the Dirac operator and $f(\lambda)$ is any continuous function that decays rapidly enough
at infinity \cite{Luscher:2004fu}.

We will work with twisted mass fermions \cite{Frezzotti:2000nk,Frezzotti:2003xj}.
Therefore, we need an expression for the topological susceptibility using this formulation.
We introduce doublets of quarks $\chi_i=(u_i\;\;d_i)^T$, where the subscript
labels the doublet. 
An example expression for the topological susceptibility is:
\begin{equation}
\label{chitop-tm}
\chi_{top}=\mu^6\; \sigma_{2;1}\equiv\frac{\langle Q^2\rangle}{V}\,,
\end{equation}
where: 
\begin{equation}
\sigma_{2;1}(\mu) = a^{20}\sum_{x_1\dots
x_5}\langle S^+_{41}(x_1)\;P^-_{12}(x_2)\;P^+_{23}(x_3)\;P^-_{34}(x_4)\;
S^+_{56}(x_5)\;P^-_{65}(0)\rangle\,,
\end{equation}
and $S^\pm_{ij}=\chibar_{i}\tau^\pm\chi_{j}$, 
$P^\pm_{ij}=\chibar_{i}\tau^\pm\gamma_5\chi_{j}$,
$V$ is the volume
and all twisted quark masses are taken to be $\mu$.
It can be shown that this definition of $\chi_{top}$ is related to
the following spectral sum:
\begin{equation}
 \sigma_{k;l}(\mu) = \left\langle \Tr \left\{\gamma_5(D^\dagger D+\mu^2)^{-k}\right\}
\Tr \left\{\gamma_5(D^\dagger D+\mu^2)^{-l}\right\} \right\rangle
\end{equation} 
and hence its computation can be
carried out with spectral projectors $\PM$, which project into the subspace spanned by the
eigenvectors of $D^\dagger D$ that correspond to all eigenvalues that are below some threshold value
$M^2$. For details of spectral projectors, we refer to the original publication \cite{Giusti:2008vb}.

Since twisted mass fermions are not chirally invariant, Eq.~\eqref{chitop-tm} for $\chi_{top}$ is not
renormalization group invariant.
The relevant multiplicative renormalizations are the following: $P_{R}=Z_P P$, $S_{R}=Z_S S$,
$\mu_{R}={Z_P}^{-1}\mu$, where the subscript $R$ denotes renormalized quantities.
The presence of two scalar densities in Eq.~\eqref{chitop-tm} hence implies the renormalization of
$\chi_{top}$ with $(Z_S/Z_P)^2$, i.e. the expression for
the renormalized susceptibility, $\chi_{top,R}$, reads:
\begin{equation}
\label{eq:chi}
\chi_{top,R} = \frac{Z_S^2}{Z_P^2} \frac{\langle Q^2\rangle}{V}. 
\end{equation} 
In the following, we will drop the subscript $R$ and always consider the renormalized topological
susceptibility.
The evaluation of this observable with spectral projectors is straightforward and for the details we
refer to Ref.~\cite{Luscher:2010ik}.
Here we just give the final formulae:
\begin{equation}
\label{eq:chi2}
\chi_{top}=\frac{Z_S^2}{Z_P^2}
  \frac{\langle{\cal C}^2\rangle-\frac{\langle{\cal B}\rangle}{N}}{V}, 
\end{equation} 
where $N$ is the number of used stochastic sources,
\begin{equation}
  {\cal C}={1\over N}\sum_{k=1}^N
  \left(\RM\eta_k,\dirac{5}\RM\eta_k\right),
 \end{equation}
\begin{equation}
\label{eq:B}
 {\cal B}={1\over N}\sum_{k=1}^N
  \left(\RM\dirac{5}\RM\eta_k,\RM\dirac{5}\RM\eta_k\right),
\end{equation}
where $\RM$ is a rational approximation to the projector $\PM$ and $\eta_k$ are randomly generated
pseudofermion fields added to the theory.
Let us note that in the limit of an infinite number
of stochastic sources, the observable ${\cal C}$ can be associated with the topological
charge (cf. Eqs.~\eqref{eq:chi} and \eqref{eq:chi2}), whose distribution,
however, has to be corrected for the finite number of stochastic sources $N$ to obtain the correct
topological susceptibility.

\section{$O(a)$-improvement with twisted mass fermions at maximal twist}
In this section, we sketch the proof of $O(a)$-improvement of the topological susceptibility
evaluated
with twisted mass fermions at maximal twist.
For more details, we refer to an upcoming publication \cite{letter}.

 For an on-shell observable to be automatically improved, it
is sufficient that it remains invariant under the $\mathcal{R}_5^{1,2}$ transformations defined by:
$\chi_i(x)\rightarrow i\gamma_5\tau^{1,2}\chi_i(x)$, $\overline{\chi}_i(x)\rightarrow \overline{\chi}_i(x)
i\gamma_5\tau^{1,2}$, where with the subscript $i$ we generically indicate valence and sea twisted mass
doublets.

As an example of a definition of the topological susceptibility in terms of density
chains we consider Eq.~\eqref{chitop-tm} 
where we take $Q^2$ expressed in terms of two closed density chains -- one with 4 densities and the
other with only 2, such that the total number is not smaller than 5, to guarantee the absence of
non-integrable short-distance singularities.
One can verify that $\chi_{top}$ given by the above formula is
$\mathcal{R}_5^{1,2}$-parity even up to a charge conjugation transformation.
However, the automatic $O(a)$-improvement can still be spoiled by contact terms.
In the following, we will show that appropriate combinations of such contact terms are
$\mathcal{R}_5^{1,2}$-parity odd and
hence automatic $O(a)$-improvement is preserved at maximal twist.

We start with the Symanzik effective theory expansion of the renormalized $\sigma_{2;1}$:
\begin{align}
\label{Symanzik}
&\sigma_{2;1,R}=\int d^4x_1\dots
d^4x_5\expect{S^+_{41}(x_1)P^-_{12}(x_2)P^+_{23}(x_3)P^-_{34}(x_4)S^+_{56}(x_5)P^-_{65}(0)}_0\nonumber\\
&+a~{\rm S.T.} + a~{\rm C.T}+O(a^2)\,,
\end{align}
where the densities on the right-hand side are renormalized operators that with abuse of notation
we denote as the lattice densities. 
The term labelled with S.T. corresponds to the standard terms
appearing in the Symanzik expansion and the one labelled with C.T.
corresponds to the $O(a)$ terms arising from the short distance singularities in the 
product of two densities. 
If we tune our lattice action parameters to achieve maximal
twist, one can use the standard arguments leading to automatic $O(a)$ improvement to show that the
S.T. vanish.
An example of the C.T. is given by
\begin{align}
\label{eq:ex_ct}
&\int d^4x_2 d^4x_3
d^4x_4 d^4x_5\expect{P^\uparrow_{42}(x_2)P^+_{23}(x_3)P^-_{34}(x_4)S^+_{56}(x_5)P^-_{65}(0)}_0
\nonumber\\
&+\int d^4x_1 d^4x_2 d^4x_3
d^4x_5\expect{P^\downarrow_{31}(x_1)P^-_{12}(x_2)P^+_{23}(x_3)S^+_{56}(x_5)P^-_{65}(0)}_0\,,
\end{align}
where the subindex $\expect{}_0$ denotes continuum expectation values and we have introduced the
short-hand notation
$P^{\uparrow,\downarrow}_{ij}=\chibar_{i}\left(\frac{\mathbbm{1}\pm\tau^3}{2}\right)\gamma_5\chi_{j}$. 
These additional $O(a)$ terms originate from applying OPE to all pairs of consecutive densities 
to the lattice correlator, which corresponds to
the contact terms that can introduce $O(a)$ effects\footnote{Three or
more densities at the same point lead to cut-off effects of $O(a^n)$ with $n\geq2$.}.
For example the product of a scalar and a pseudoscalar density at short distance 
receives several contributions and the one with the smallest dimension is proportional 
to the pseudoscalar density. 

It is clear from the Symanzik expansion that for $O(a)$-improvement to be maintained, the terms labelled as 
C.T. on the right-hand side (RHS) of Eq.~\eqref{Symanzik}
have to vanish. 

As an example let us consider the term in Eq.~\eqref{eq:ex_ct}. If we perform
an $\mathcal{R}_5^{1}$ transformation only for doublets $1-4$ we obtain
\begin{align}
\label{obs1}
\expect{P^\uparrow_{42}P^+_{23}P^-_{34}S^+_{56}P^-_{65}}_0
+\expect{P^\downarrow_{31}P^-_{12}P^+_{23}S^+_{56}P^-_{65}}_0
\xrightarrow{\mathcal{R}_5^1}
-\expect{P^\downarrow_{42}P^-_{23}P^+_{34}S^+_{56}P^-_{65}}_0
-\expect{P^\uparrow_{31}P^+_{12}P^-_{23}S^+_{56}P^-_{65}}_0\,.
\end{align}
Up to a relabelling of flavors this linear combination is odd under $\mathcal{R}_5^{1,2}$, 
i.e. it vanishes for twisted mass fermions at maximal twist.

It can be similarly shown, by grouping the other terms in C.T. of Eq.~\eqref{Symanzik},
that all terms that appear in the Symanzik expansion are $\mathcal{R}_5^{1,2}$
odd, up to a charge conjugation transformation,
and hence vanish in the continuum limit and do not introduce additional $O(a)$ cutoff effects. 
Moreover, this proof holds also in the general case -- for any
density chain that can be written in terms of $D^\dagger D$ (i.e. containing an even number of
pseudoscalar and scalar densities).

\section{Numerical results}
In the previous section, we have shown that the topological susceptibility computed with spectral
projectors is automatically $O(a)$-improved at maximal twist.
Now, we will present a summary of our numerical results.
For the setup and details of the simulations, we refer to
Refs.~\cite{Baron:2010bv,Boucaud:2008xu,Cichy:2013chi,Cichy:2013gja}.

\begin{figure}[]
\subfigure[$N_f=2$, $\beta=3.9$]{
\includegraphics[height=.345\textwidth]{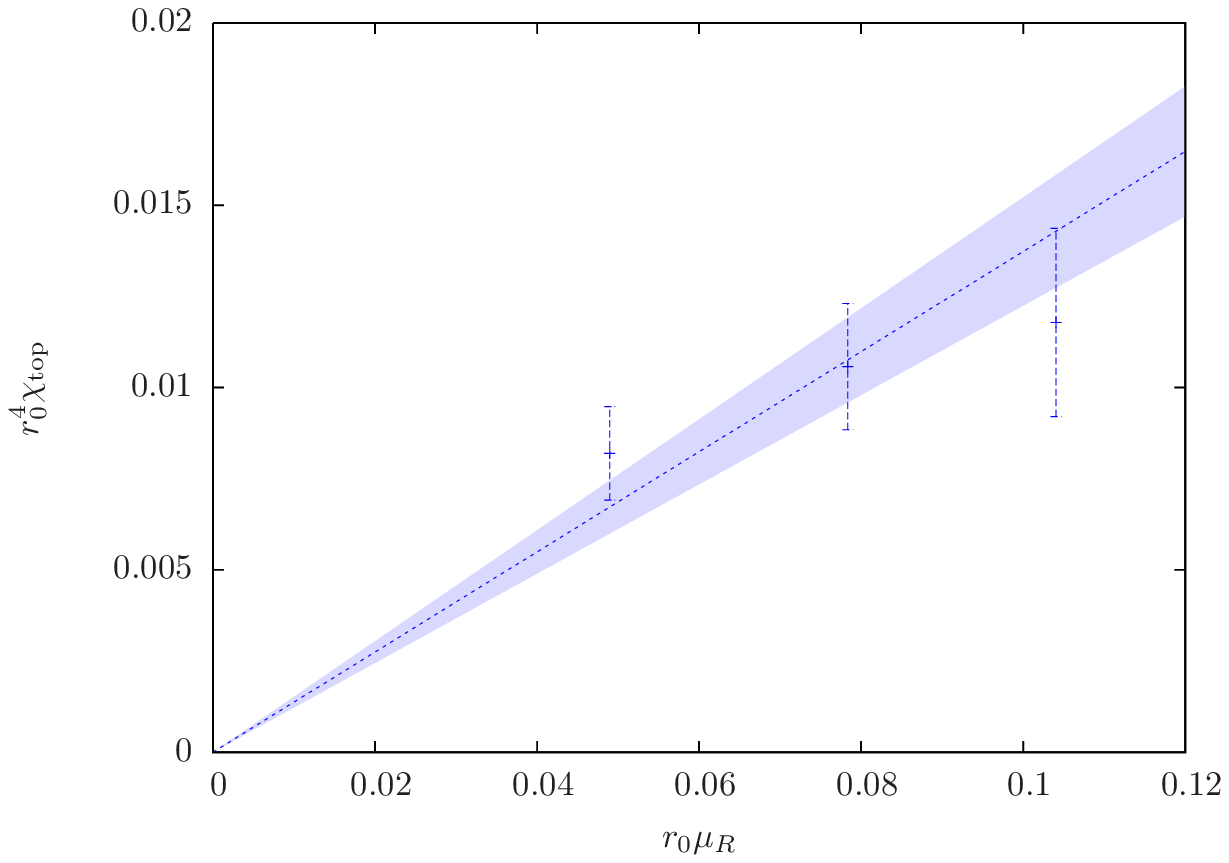}
}
\subfigure[$N_f=2+1+1$, $\beta=1.9$]{
\includegraphics[height=.345\textwidth]{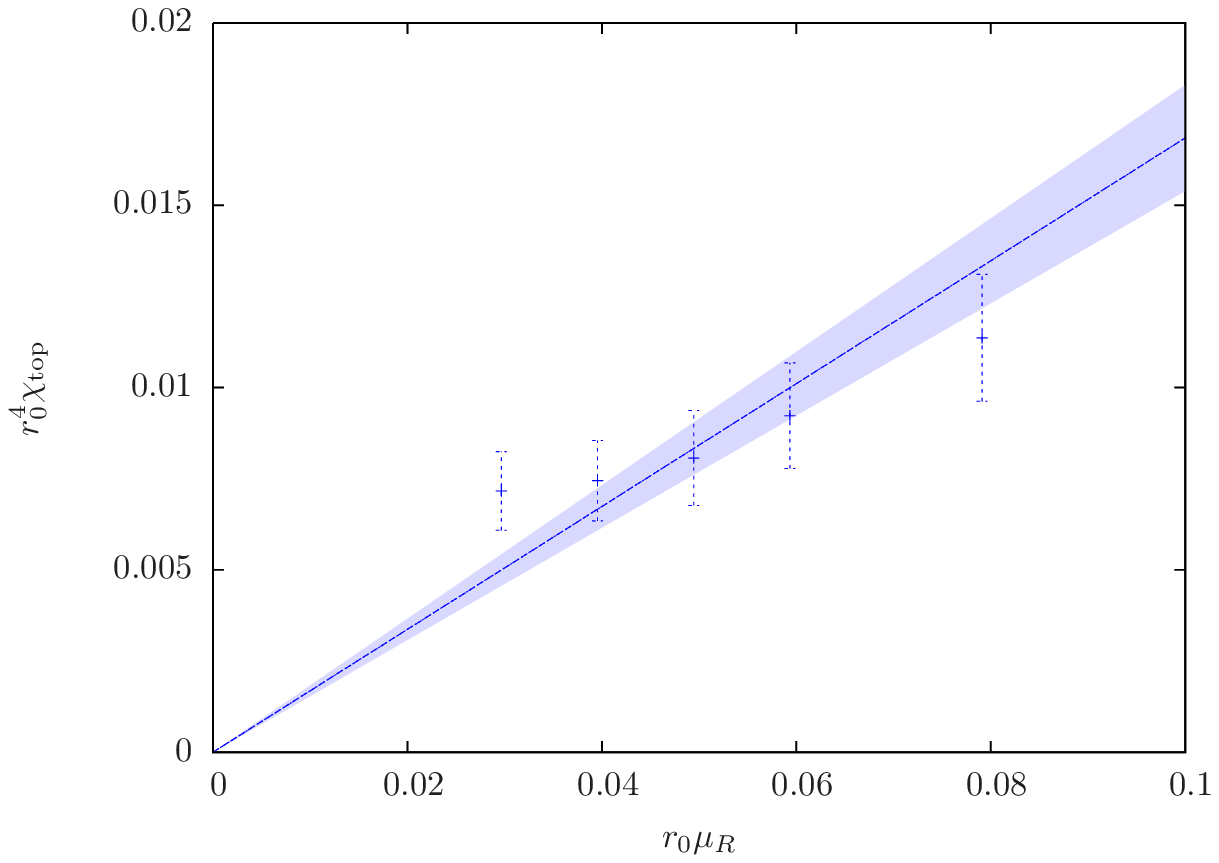}
}
\subfigure[$N_f=2+1+1$, $\beta=1.95$]{
\includegraphics[height=.345\textwidth]{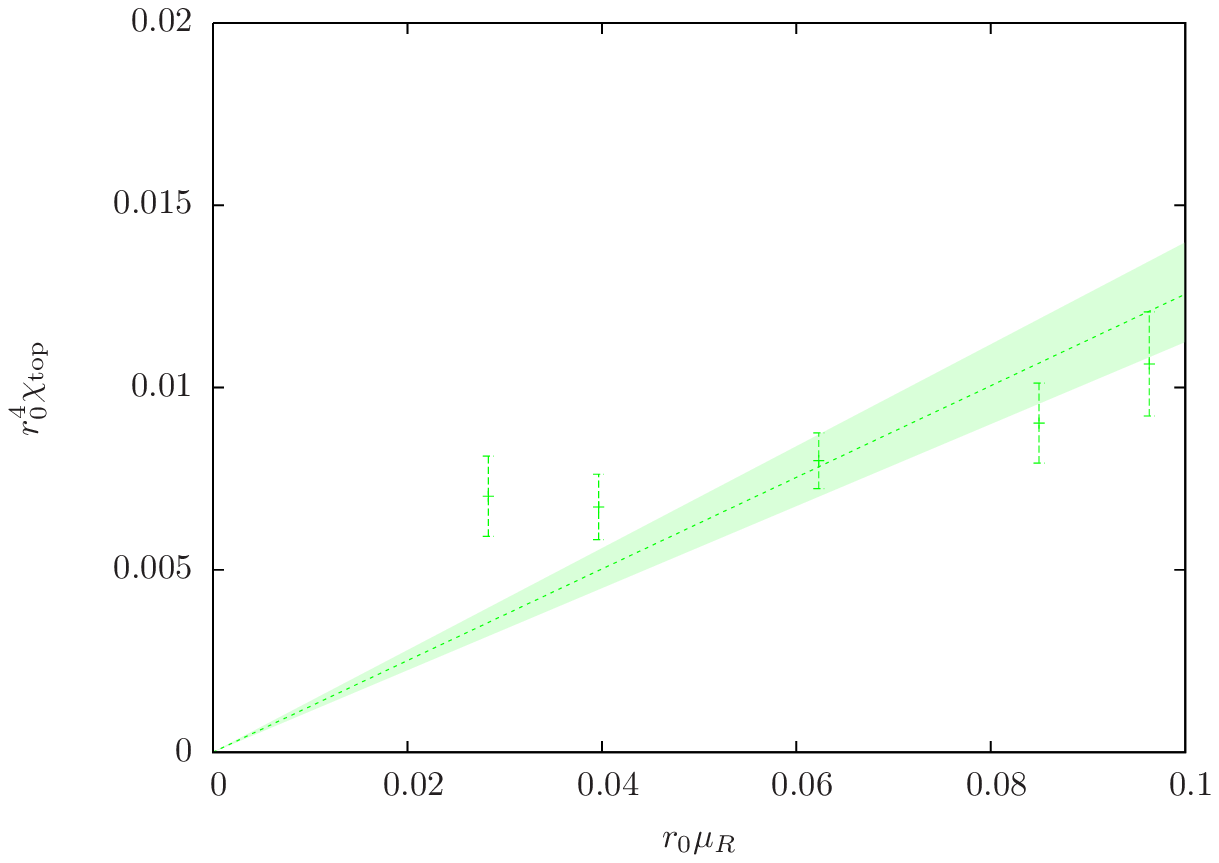}
}
\subfigure[$N_f=2+1+1$, $\beta=2.1$]{
\includegraphics[height=.345\textwidth]{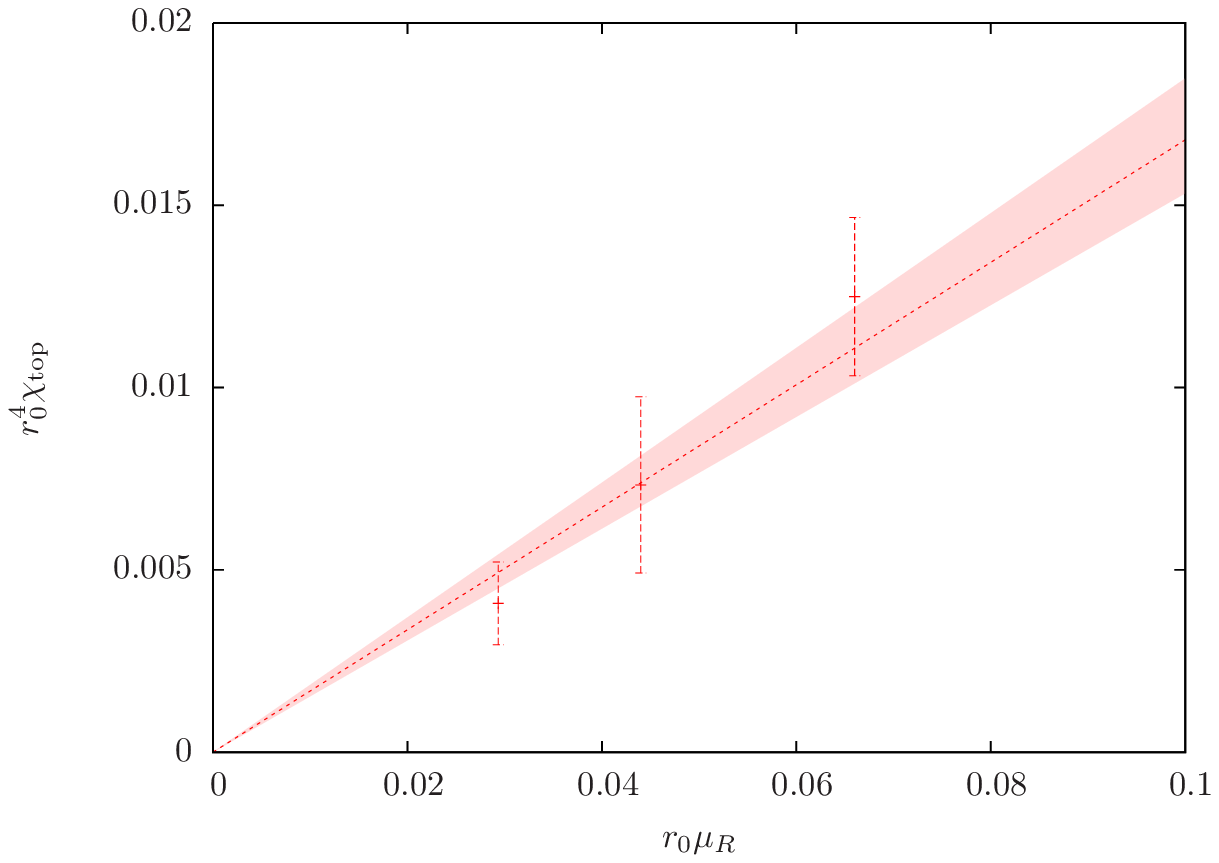}
}
\caption{Renormalized quark mass dependence of the topological susceptibility for
$N_f=2$ (a) and $N_f=2+1+1$ (b,c,d). The straight line corresponds to a fit of LO SU(2) $\chi$PT.}
\label{fig:chiral}
\end{figure}

\subsection{$N_f=2$}
Our 2-flavour results are at a single lattice spacing of approx. 0.085 fm ($\beta=3.9$), with pion
masses ranging from 300 to 450 MeV.
The quark mass dependence of the topological susceptibility is shown in Fig.~\ref{fig:chiral}(a).
Within the rather large errors, the behaviour of $\chi_{top}$ is compatible with LO$\chi$PT:
$\chi_{top}=\Sigma\mu/N_f$,
where $\Sigma$ is the chiral condensate and $N_f$ the number of light flavours.
The fit gives $r_0\Sigma^{1/3}=0.650(22)$ ($\MSb$ scheme at $\mu=2$ GeV).
The quoted error includes the statistical error and uncertainties from $r_0/a$ and $Z_P/Z_S$
determinations.
This can be compared to our results from direct extraction: 0.696(20) (at $\beta=3.9$ in the chiral
limit) or 0.689(33) (in the continuum limit and in the chiral limit) \cite{Cichy:2013gja}.

\begin{figure}[]
\begin{minipage}[c]{0.45\textwidth}
\includegraphics[height=.72\textwidth]{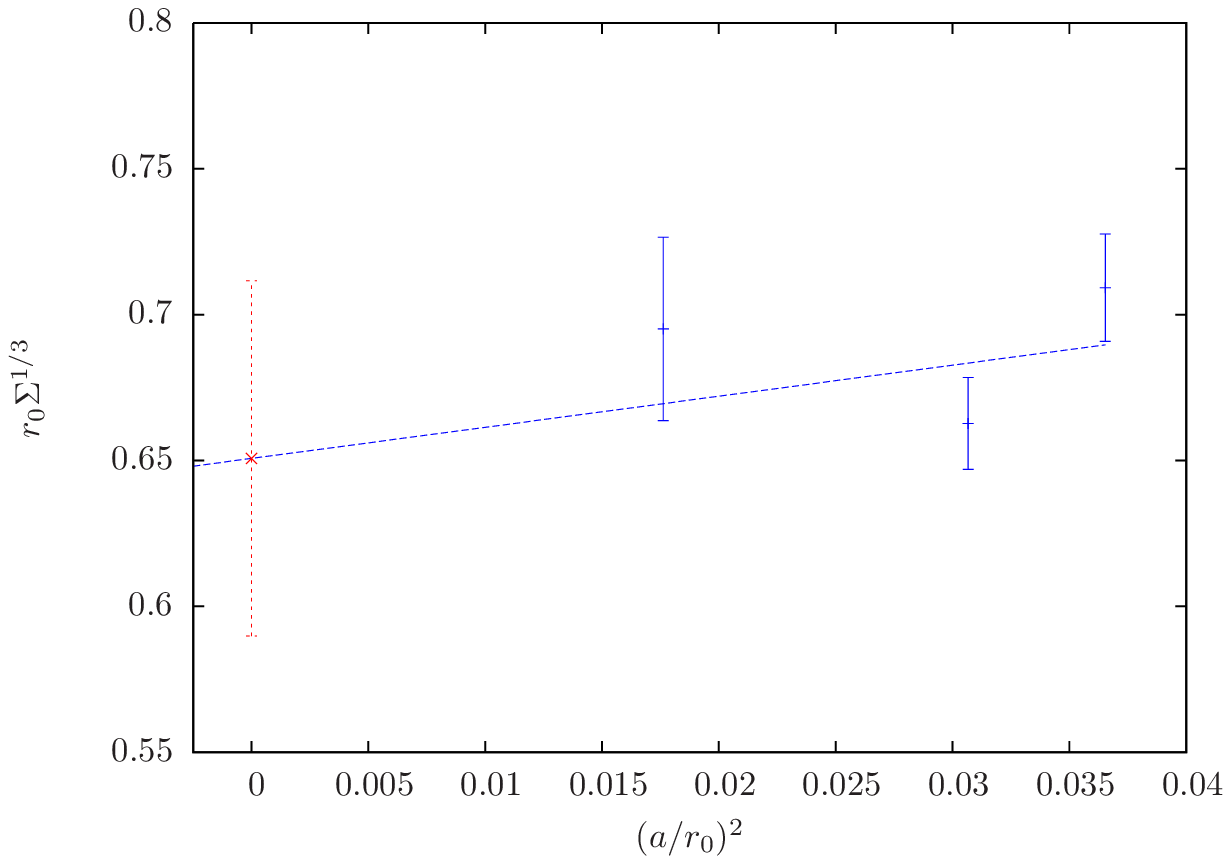}
\caption{Continuum limit of the chiral condensate $r_0\Sigma^{1/3}$ ($\MSb$ scheme at $\mu=2$ GeV)
extracted from $\chi_{top}$.}
\label{fig:cont}
\end{minipage}
\hspace*{0.1\textwidth}
\begin{minipage}[c]{0.45\textwidth}
\includegraphics[height=.72\textwidth]{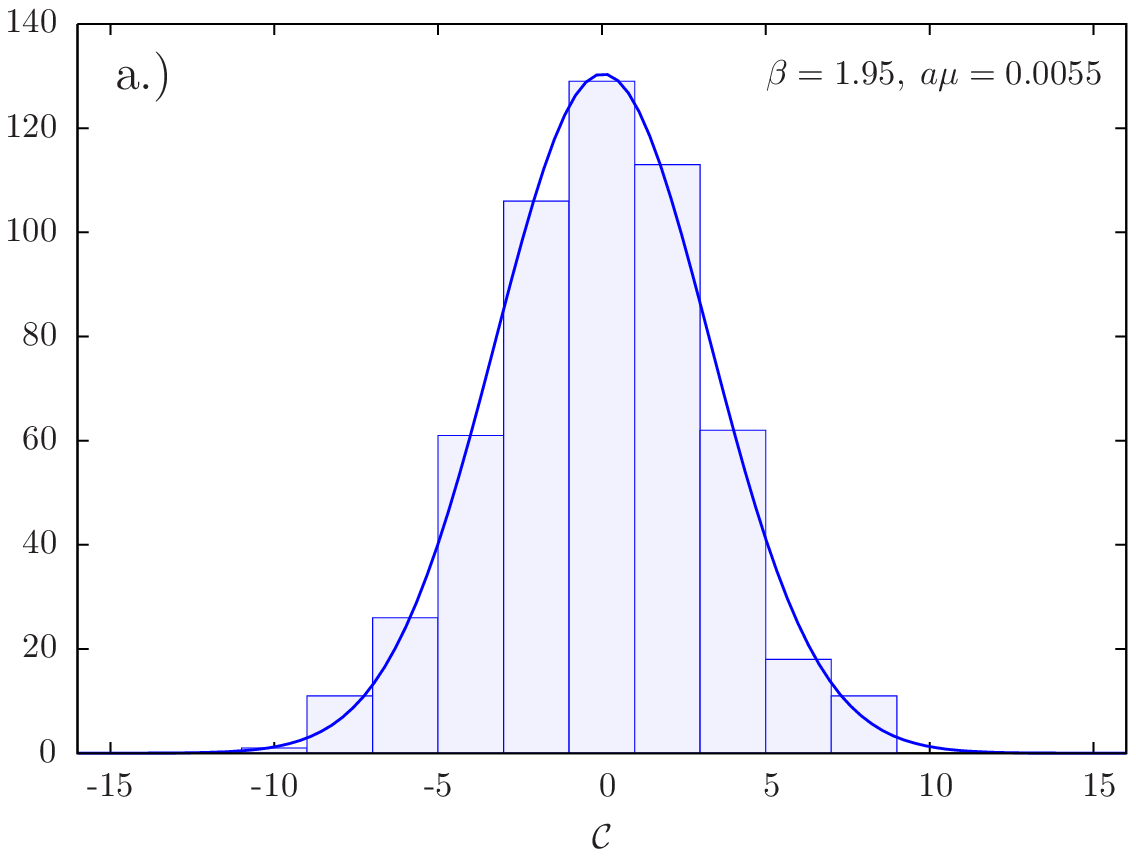}
\caption{Histogram of the observable ${\cal C}$ for the ensemble: $N_f=2+1+1$, $\beta=1.95$,
$a\mu=0.0055$.}
\label{fig:histo}
\end{minipage}
\end{figure}

\subsection{$N_f=2+1+1$}
In the case of $N_f=2+1+1$ simulations, we present data at 3 lattice spacings ranging from around
0.061 to 0.086 fm, with pion masses down to 260 MeV. 
The chiral behaviour of the topological susceptibility is shown in Figs.~\ref{fig:chiral}(b)-(d),
separately at each lattice spacing.
At our finest lattice spacing ($\beta=2.1$), we see a tendency towards the expected suppression of the
topological susceptibility. 
The data are compatible with SU(2) LO$\chi$PT and since the data are well described by the LO
expression, we only attempted linear fits of the quark mass dependence.
We compared two kinds of fits: fits to the full quark mass dependence and fits with a cut above
$r_0\mu_R=0.07$ (around $m_\pi=400$ MeV) to account for the fact that LO$\chi$PT is not expected to work
well at heavy pion masses.
We extracted the values of $r_0\Sigma^{1/3}$ at each lattice spacing and then performed a continuum
limit extrapolation for both types of fits (in Fig.~\ref{fig:cont} we show only the fits with a mass
cut), obtaining the following value of the chiral condensate in the chiral limit ($\MSb$ scheme at
$\mu=2$ GeV):
$r_0\Sigma^{1/3}=0.651(61)$ (error as for $N_f=2$, combined with the error from continuum
extrapolation). The full fits yield in the continuum limit a compatible result:
$r_0\Sigma^{1/3}=0.619(58)$.
We find good agreement with our direct determination:
$r_0\Sigma^{1/3}=0.680(29)$ \cite{Cichy:2013gja}.

To conclude this proceeding, we would like to emphasize one important aspect. Our typical precision for
$\chi_{top}$ at any given lattice spacing is of the order of 15-20\%. 
Such precision is not enough for robust NLO$\chi$PT fits and hence we only performed linear LO fits.
However, for one of our ensembles ($N_f=2+1+1$, $\beta=1.95$, $a\mu=0.0055$), we have considerably
better statistics (a factor of nearly 4 higher than typical for other ensembles).
This allows to obtain a 9\% precision in $\chi_{top}$ and sample all relevant topological sectors
correctly -- this is shown in the histogram of the observable ${\cal C}$ (Fig.~\ref{fig:histo}), which
is almost perfectly Gaussian.
Unfortunately, for other ensembles we do not achieve good, i.e. symmetric and centered around zero
Gaussian histograms (although they are compatible with Gaussian within rather large errors) and we are
not able to increase our statistics within the existing ensembles due to autocorrelations.
This means that a precise analysis of the topological susceptibility (i.e. with errors $\leq10\%$)
requires considerably longer Monte Carlo runs than typically generated for most Lattice QCD applications.

\section*{Acknowledgments}
We thank the European Twisted Mass Collaboration (ETMC) for generating gauge field ensembles used in this
work and several useful discussions.
K.C. was supported by Foundation for Polish Science fellowship ``Kolumb''.
This work was supported in part by the DFG Sonderforschungsbereich/Transregio SFB/TR9. 
K.J. was supported in part by the Cyprus Research Promotion
Foundation under contract $\Pi$PO$\Sigma$E$\Lambda$KY$\Sigma$H/EM$\Pi$EIPO$\Sigma$/0311/16.
The computer time for this project was made available to us by the J\"ulich
Supercomputing Center, LRZ in Munich, the PC cluster in Zeuthen, Poznan Supercomputing and Networking
Center (PCSS).
We also would like to thank L. Giusti and M. L\"uscher for insightful discussions.

\bibliography{lat13-chi}

\providecommand{\href}[2]{#2}\begingroup\raggedright\begin{thebibliography}{10}

\bibitem{Giusti:2004qd}
L.~Giusti, G.~Rossi, and M.~Testa, {\it {Topological susceptibility in full QCD
  with Ginsparg-Wilson fermions}},  {\em Phys.Lett.} {\bf B587} (2004)
  157--166, [\href{http://xxx.lanl.gov/abs/hep-lat/0402027}{{\tt
  hep-lat/0402027}}].

\bibitem{Luscher:2004fu}
M.~Luscher, {\it {Topological effects in QCD and the problem of short distance
  singularities}},  {\em Phys.Lett.} {\bf B593} (2004) 296--301,
  [\href{http://xxx.lanl.gov/abs/hep-th/0404034}{{\tt hep-th/0404034}}].

\bibitem{Giusti:2008vb}
L.~Giusti and M.~Luscher, {\it {Chiral symmetry breaking and the Banks-Casher
  relation in lattice QCD with Wilson quarks}},  {\em JHEP} {\bf 0903} (2009)
  013, [\href{http://xxx.lanl.gov/abs/0812.3638}{{\tt arXiv:0812.3638}}].

\bibitem{Luscher:2010ik}
M.~Luscher and F.~Palombi, {\it {Universality of the topological susceptibility
  in the SU(3) gauge theory}},  {\em JHEP} {\bf 1009} (2010) 110,
  [\href{http://xxx.lanl.gov/abs/1008.0732}{{\tt arXiv:1008.0732}}].

\bibitem{Cichy:2011an}
K.~Cichy, V.~Drach, E.~Garcia-Ramos, and K.~Jansen, {\it {Topological
  susceptibility and chiral condensate with $N_f=2+1+1$ dynamical flavors of
  maximally twisted mass fermions}},  {\em PoS} {\bf LATTICE2011} (2011) 102,
  [\href{http://xxx.lanl.gov/abs/1111.3322}{{\tt arXiv:1111.3322}}].

\bibitem{Cichy:2013chi}
K.~Cichy, E.~Garcia-Ramos, and K.~Jansen, ``{Topological susceptibility from
  the twisted mass Dirac operator spectrum}.'' in preparation.

\bibitem{Frezzotti:2003ni}
R.~Frezzotti and G.~Rossi, {\it {Chirally improving Wilson fermions. 1. O(a)
  improvement}},  {\em JHEP} {\bf 0408} (2004) 007,
  [\href{http://xxx.lanl.gov/abs/hep-lat/0306014}{{\tt hep-lat/0306014}}].

\bibitem{Cichy:2013egr}
K.~Cichy, E.~Garcia-Ramos, K.~Jansen, and A.~Shindler, {\it {Computation of the
  chiral condensate using $N_f = 2$ and $N_f = 2 + 1 + 1$ dynamical flavors of
  twisted mass fermions}},  {\em PoS} {\bf LATTICE2013} (2013) 128.

\bibitem{Frezzotti:2000nk}
{\bf Alpha} Collaboration, R.~Frezzotti, P.~A. Grassi, S.~Sint, and P.~Weisz,
  {\it {Lattice QCD with a chirally twisted mass term}},  {\em JHEP} {\bf 0108}
  (2001) 058, [\href{http://xxx.lanl.gov/abs/hep-lat/0101001}{{\tt
  hep-lat/0101001}}].

\bibitem{Frezzotti:2003xj}
R.~Frezzotti and G.~Rossi, {\it {Twisted mass lattice QCD with mass
  nondegenerate quarks}},  {\em Nucl.Phys.Proc.Suppl.} {\bf 128} (2004)
  193--202, [\href{http://xxx.lanl.gov/abs/hep-lat/0311008}{{\tt
  hep-lat/0311008}}].

\bibitem{letter}
K.~Cichy, E.~Garcia-Ramos, K.~Jansen, and A.~Shindler in preparation.

\bibitem{Baron:2010bv}
{\bf ETMC} Collaboration, R.~Baron et~al., {\it {Light hadrons from lattice QCD
  with light (u,d), strange and charm dynamical quarks}},  {\em JHEP} {\bf
  1006} (2010) 111, [\href{http://xxx.lanl.gov/abs/1004.5284}{{\tt
  arXiv:1004.5284}}].

\bibitem{Boucaud:2008xu}
{\bf ETMC} Collaboration, P.~Boucaud et~al., {\it {Dynamical Twisted Mass
  Fermions with Light Quarks: Simulation and Analysis Details}},  {\em
  Comput.Phys.Commun.} {\bf 179} (2008) 695--715,
  [\href{http://xxx.lanl.gov/abs/0803.0224}{{\tt arXiv:0803.0224}}].

\bibitem{Cichy:2013gja}
K.~Cichy, E.~Garcia-Ramos, and K.~Jansen, {\it {Chiral condensate from the
  twisted mass Dirac operator spectrum}},  {\em JHEP} {\bf 1310} (2013) 175,
  [\href{http://xxx.lanl.gov/abs/1303.1954}{{\tt arXiv:1303.1954}}].

\end{thebibliography}\endgroup

\end{document}